\pdfoutput=1
\documentclass[letter,twocolumn]{jpsj3}
\usepackage{txfonts}
\usepackage{graphicx,color}

\title{Thermodynamic Investigation of Metamagnetism in Pulsed High Magnetic Fields on Heavy Fermion Superconductor UTe$_2$}

\author{
Shusaku Imajo$^1$\thanks{imajo@issp.u-tokyo.ac.jp}, 
Yoshimitsu Kohama$^1$,
Atsushi Miyake$^1$,
Chao Dong$^1$,
Masashi Tokunaga$^1$,\\
Jacques Flouquet$^2$, 
Koichi Kindo$^1$, and 
Dai Aoki$^{2,3}$
}

\inst{
$^1$The Institute for Solid State Physics, University of Tokyo, Kashiwa, Chiba 277-8581, Japan\\
$^2$University Grenoble Alpes, CEA, IRIG-PHELIQS, F-3800 Grenoble, France\\
$^3$Institute for Materials Research, Tohoku University, Oarai, Ibaraki 311-1313, Japan \\
} 

\abst{
We investigated the thermodynamic property of the heavy fermion superconductor UTe$_2$ in pulsed high magnetic fields.
The superconducting transition in zero field was observed at $T_{\rm c}$=1.65~K as a sharp heat capacity jump.
Magnetocaloric effect measurements in pulsed-magnetic fields obviously detected a thermodynamic anomaly accompanied by a first-order metamagnetic transition at $\mu$$_{0}$$H_{\rm m}$=36.0~T when the fields are applied nearly along the hard-magnetization $b$-axis.
From the results of heat capacity measurements in magnetic fields, we found a drastic diverging electronic heat capacity coefficient of the normal state $\gamma$$_{\rm N}$ with approaching $H_{\rm m}$.
Comparing with the previous works via the magnetic Clausius-Clapeyron relation, we unveil the thermodynamic details of the metamagnetic transition.
The enhancement of the effective mass observed as the development of $\gamma_{\rm N}$ indicates that quantum fluctuation strongly evolves around $H_{\rm m}$; it assists the superconductivity emerging even in extremely high fields.}

\begin{document}
\maketitle

The occurrence of unconventional triplet superconductivity (SC) in ferromagnetic (FM) compounds begins with the discovery of superconductivity in UGe$_2$~\cite{Sax00} with the singularity
that SC emerges just at the crossing between two FM phases (FM2 and FM1) at a pressure $P_{\rm x}\sim 1.2\,{\rm GPa}$;
it collapses on entering in the paramagnetic (PM) ground state above $P_{\rm c}\sim 1.6\,{\rm GPa}$.
Lowering the Curie temperature 
in UGe$_2$ ($T_{\rm Curie}=52\,{\rm K}$) and
URhGe ($T_{\rm Curie}=9.5\,{\rm K}$)
leads to the opportunity to modify the ferromagnetic interactions by the magnetic field~\cite{Aok19,Aok12_JPSJ_review}.
A spectacular phenomenon observed in URhGe~\cite{Aok01} is that transverse magnetic field ($H\parallel b$, hard magnetization axis) with respect to the ferromagnetic sublattice magnetization ($M\sim 0.4\,\mu_{\rm B}$, $H\parallel c$-axis, easy-magnetization axis)
pushes $T_{\rm Curie}$ to zero at a spin-reorientation field $\mu$$_{0}$$H_{\rm R}\sim 12\,{\rm T}$
and the field-reentrant superconductivity (RSC) appears between $8$ to $13\,{\rm T}$~\cite{Lev05,Lev07}.
The origin of RSC is the field enhancement of the effective mass $m^\ast$~\cite{Miy08,Miy09}, which may result from the combined effects of ferromagnetic fluctuations and Fermi surface instabilities associated with the Lifshitz transition~\cite{She18}.
Metamagnetic transition in itinerant ferromagnetic materials is often connected with
the so called wing structure of PM--FM boundary close to the critical pressure $P_{\rm c}$ in the phase diagram.
In uranium compounds, well-known cases are UGe$_2$~\cite{Tau10,Kot11} and UCoAl~\cite{Aok11_UCoAl,Kim15},
when the field is applied along the Ising easy-magnetization axis.
As it is known in URhGe, a new key phenomenon in UTe$_2$ is the field evolution of the magneto-crystalline energy.

The new feature with the discovery of SC in UTe$_2$ is that
the ground state in the normal state is paramagnetic at the verge of FM order~\cite{Ran2018,Aoki2019,Sundar2019}.
When the field is applied along the $b$-axis in the orthorhombic structure,
a sharp metamagnetic transition occurs at $\mu$$_{0}$$H_{\rm m}=35\,{\rm T}$~\cite{Miyake2019,Ran2019,Knafo2019},
leading to a polarized ferromagnetic phase with a huge jump $\Delta M\sim 0.6\,\mu_{\rm B}$ of the magnetization~\cite{Miyake2019}, which is six times larger than that of URhGe ($\Delta M\sim 0.1\,\mu_{\rm B}$)~\cite{Lev05}.
In UTe$_2$, at low field, the easy-magnetization axis is now $a$-axis; above $H_{\rm m}$ the $b$-axis becomes the easy-magnetization axis.
The strong enhancement of SC is observed as the field increases towards $H_{\rm m}$~\cite{Knebel2019,Ran2019}.
Our aim is to present direct heat capacity measurements down to $0.8\,{\rm K}$ at lower temperature, comparing with the previous estimation of the field variation of $m^\ast$.
We focus on the field dependence of the linear temperature term of the heat capacity ($\gamma_{\rm N}$), which is proportional to the effective mass $m^\ast$,
that is, $m^\ast = m^{\ast\ast} + m_{\rm b}$,
where $m_{\rm b}$ is the band mass and $m^{\ast\ast}$ is the correlation mass driven by ferromagnetic fluctuations but also by additional Fermi surface instabilities at $H_{\rm m}$.
The coupling constant of the Cooper pair is given by $\lambda = m^{\ast\ast}/m_{\rm b}$~\cite{Aok19}.

A single crystal of UTe$_2$ was grown using chemical vapor transport method~\cite{Ran2018,Aoki2019}.
Measurements of heat capacity on the single crystal of UTe$_2$, weighing 444.6 $\mu$g, were performed by the quasi-adiabatic method~\cite{Kohama2013,Kohama2019,Jiao2019} in highly stabilized magnetic fields~\cite{Kohama2015} generated by a long pulse magnet at the Internatonal MegaGauss Science Laboratory of the Institute for Solid State Physics of the University of Tokyo.
The details of the heat capacity measurements are described in the supplemental materials~\cite{suppl}.
The $b$-axis of the sample was parallel to the direction of the applied field within an accuracy of a few degrees.
Measurements of magnetocaloric effect on the same single crystal were also performed by using the same calorimeter.
This measurement can be regarded as both nearly isothermal and adiabatic conditions depending on the time scale of magnetocaloric effect.
For a sudden magnetocaloric effect enough less than the external relaxation time $\tau _1$, the field dependence of temperature $T(H)$ mimics an adiabatic (isentropic) process.

\begin{figure}
\begin{center}
\includegraphics[width=\hsize,clip]{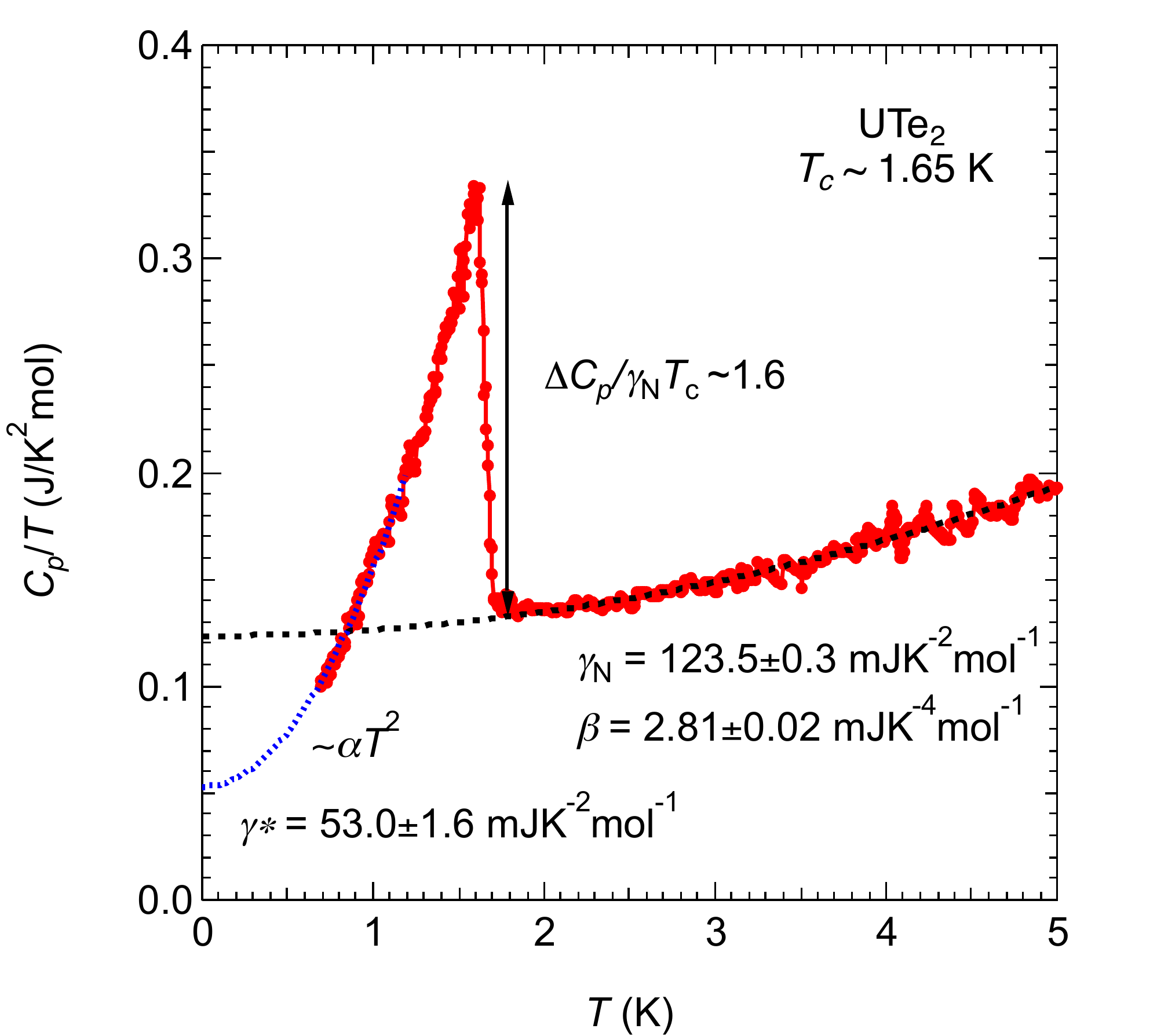}
\end{center}
\caption{(Color online) Temperature dependence of low-temperature heat capacity of a single crystal of UTe$_2$ at 0~T plotted as $C_p$/$T$ vs. $T$.
The black dotted curve indicates a least square fit to the equation, $C_p$/$T$=$\gamma _{\rm N}$+$\beta T^2$, with the electronic heat capacity coefficient of the normal state $\gamma _{\rm N}$ and the lattice heat capacity coefficient $\beta$.
The blue dotted curve represents a low-temperature extrapolation by $C_p$/$T$=$\gamma ^{\ast}$+$\alpha T^2$, where $\gamma ^{\ast}$ denotes the residual electronic heat capacity coefficient of the superconducting state and the $T^2$ term corresponds typical quasi-particle excitation expected for point-node superconductors.}
\label{fig1}
\end{figure}

 As a first characterization of the low-temperature electronic state, we report the temperature dependence of the heat capacity at 0~T in Fig.~\ref{fig1}.
The superconducting transition is clearly observed at $T_{\rm c}{\sim}$1.65~K as a sharp jump of heat capacity.
Above $T_{\rm c}$, the temperature dependence is well described by the conventional Fermi liquid behavior of the heat capacity, $C_p$/$T$=$\gamma _{\rm N}$+$\beta T^2$.
The Sommerfeld coefficient $\gamma$$_{\rm N}$ and lattice heat capacity coefficient $\beta$ are estimated as 123.5$\pm$0.3 mJK$^{-2}$mol$^{-1}$ and 2.81$\pm$0.02 mJK$^{-4}$mol$^{-1}$, respectively.
The value of $\gamma$$_{\rm N}$ agrees well with the reported values~\cite{Ran2018,Aoki2019}.
The sharp thermodynamic anomaly accompanied by the superconducting transition is also reproduced consistently with the previous works~\cite{Ran2018,Aoki2019}, which confirms the bulk superconductivity and the high quality of the present sample.
The height of the heat capacity jump at $T_{\rm c}$, $\Delta$$C_p$/$\gamma$$_{\rm N}$$T_{\rm c}$ may appear to reach about 1.6, which is slightly larger than the value, 1.43, expected for the BCS-type weak-coupling superconductivity.
At lower temperatures, the early reports succeed to represent the temperature dependence by using $C_p$/$T$=$\gamma ^{\ast}$+$\alpha T^2$, which is known as the temperature dependence of $C_p$ in the point-nodal superconductors with the residual electronic heat capacity coefficient $\gamma ^{\ast}$.
A least square fit to our data over the temperature range 0.7~K to 1.2~K yields $\gamma ^{\ast}$ = 53.0$\pm$1.6~mJK$^{-2}$mol$^{-1}$, in good agreements with the previous reports~\cite{Ran2018,Aoki2019}.
If the half of the density of states in the normal state is related with the formation of the superconductivity,
the height of the heat capacity jump at $T_{\rm c}$ gives $\Delta C_p/(\gamma _{\rm N}-\gamma ^{\ast})T_{\rm c}=2.8$;
a solid evidence for the strong-coupling superconductivity in UTe$_2$.

\begin{figure}
\begin{center}
\includegraphics[width=\hsize,clip]{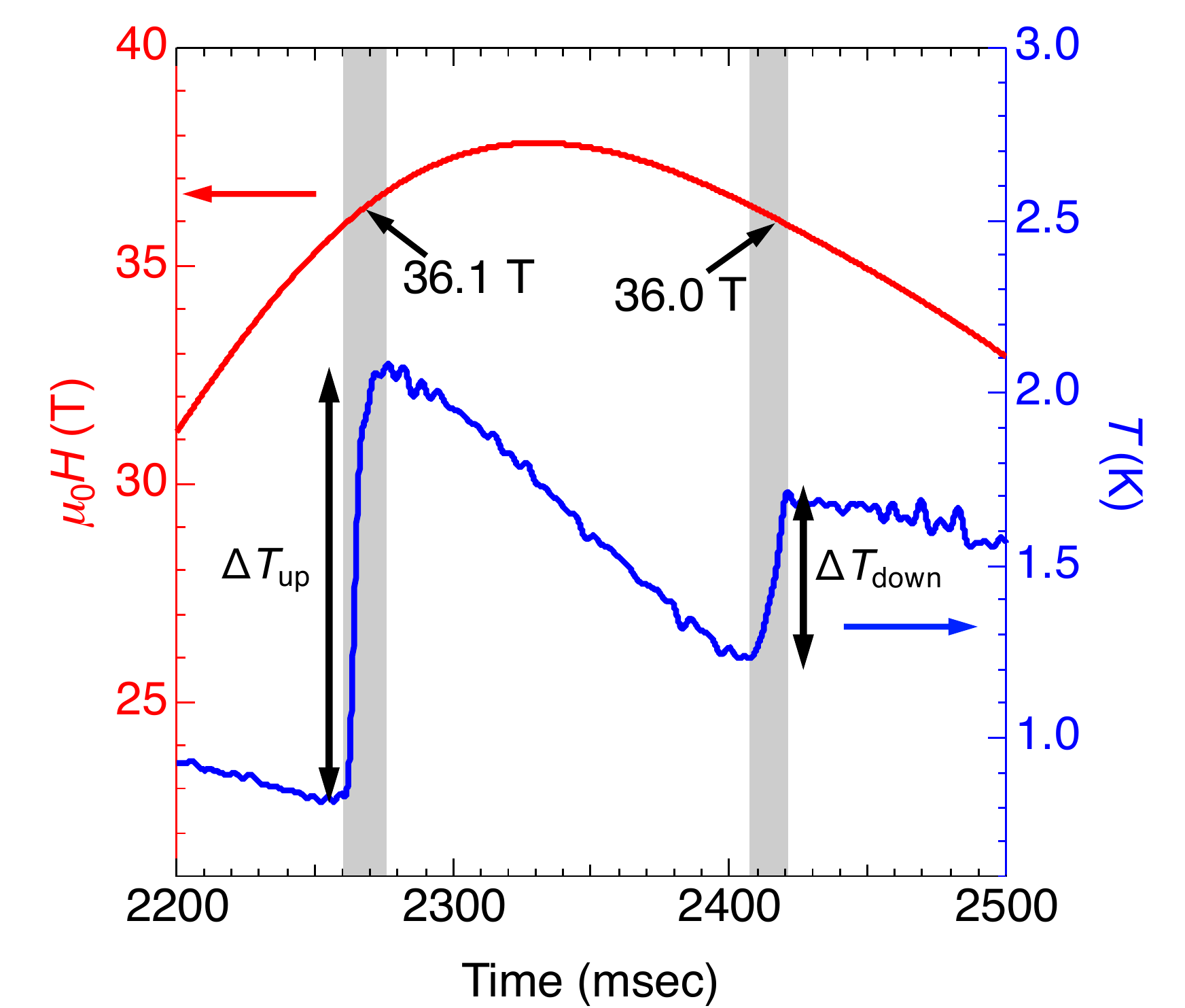}
\end{center}
\caption{(Color online) Time profiles of magnetic field and temperature in a magnetocaloric effect measurement in a pulsed-field.
At the shaded areas, the sudden heating $\Delta T_{\rm up}$ and $\Delta T_{\rm down}$ are observed when the magnetic field reaches about 36.0--36.1~T.
The time scale of the heating is $\sim$10~ms that is much less than $\tau _1$, implying that the temperature change at the phase boundary can be treated as a nearly adiabatic process.
}
\label{fig2}
\end{figure}

Next we turn our attention to the focus of this work, namely, understanding of the thermodynamic property of the metamagnetic transition.
Figure~\ref{fig2} shows the magnetocaloric effect observed in smoothly changing magnetic fields.
A sudden temperature step in the ascending part of the field pulse $\Delta T_{\rm up}$ is observed at $\sim$36.1~T, attributable to the metamagnetic transition, while the step is again detected at $\sim$36.0~T in the descending process as $\Delta T_{\rm down}$.
The sign of the temperature steps for both ascending and descending magnetic fields is positive, implying that the metamagnetic transition is considered as a first-order phase transition which release a large hysteresis loss as well as a latent heat~\cite{Silhanek2006}.
The difference between the temperature steps, $\Delta T_{\rm up}$ and $\Delta T_{\rm down}$, could be explained by the latent heat concomitant with the first-order phase transition
because the sign of the entropy change and thus the sign of heat release are opposite depending on whether the field is in the descending or ascending field pulse.
From the MCE experiment, we also notice that the present value of the critical field of 36.0--36.1~T is slightly higher than the reported values of $\mu$$_{0}$$H_{\rm m}$ ($\sim$34.9~T~\cite{Miyake2019}, $\sim$35.5~T~\cite{Knafo2019}, $\sim$35~T~\cite{Ran2019}).
The deviation of the transition field can be caused by the small misalignment of the applied magnetic fields from the crystallographic $b$-axis.
The angle dependence of $H_{\rm m}$ has been recently reported~\cite{Ran2019}.

\begin{figure}
\begin{center}
\includegraphics[width=\hsize,clip]{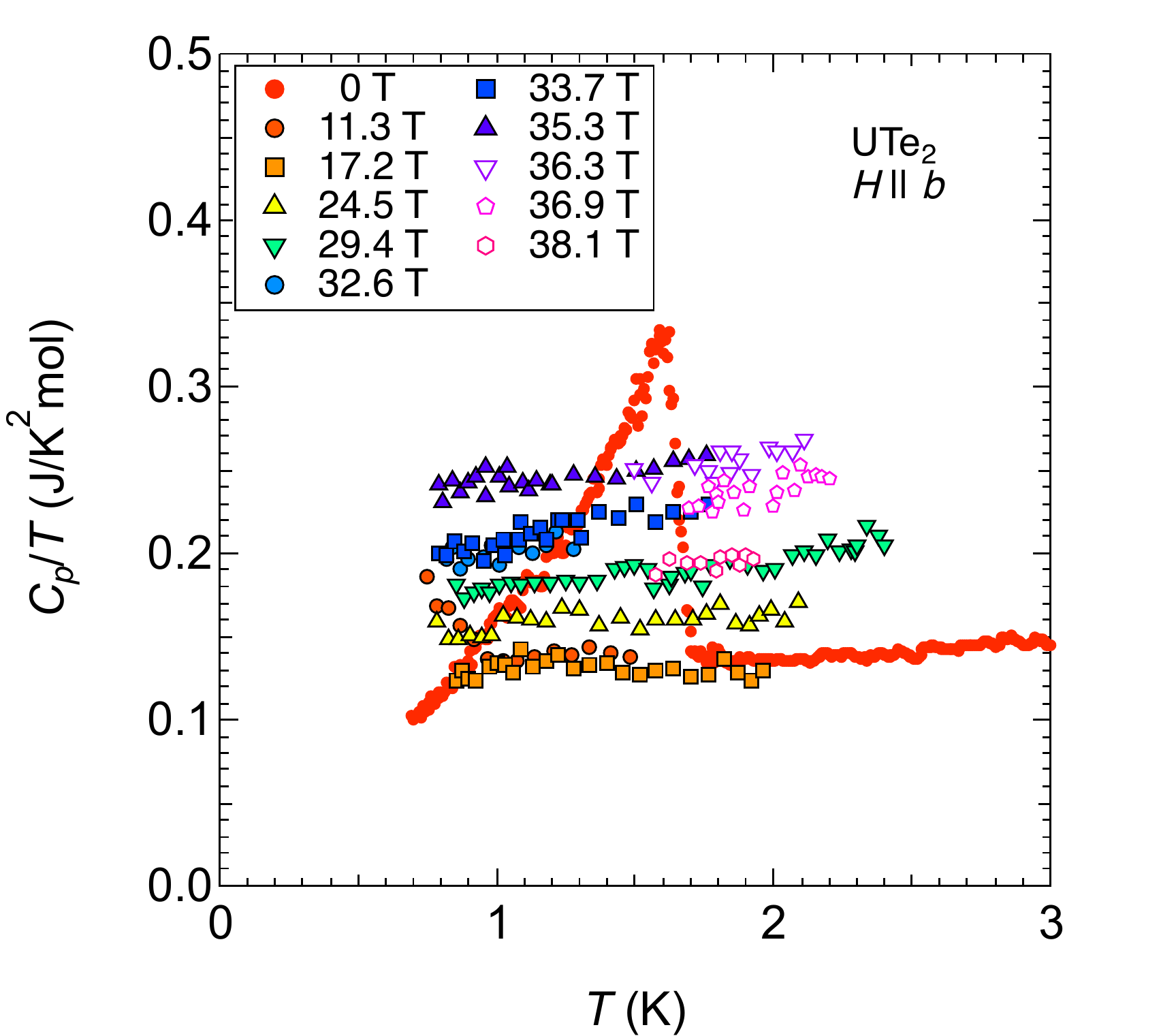}
\end{center}
\caption{(Color online) Temperature dependence of heat capacity plotted as the $C_p$/$T$ vs. $T$ at various magnetic fields.
The thermodynamic anomaly below 1~K at 11.5~T originates from the superconducting transition.
The extrapolations of the data points to zero temperature yields $\gamma$$_{\rm N}$ for each fields.
No contribution from the nuclear heat capacity was observed for the present study.}
\label{fig3}
\end{figure}

\begin{figure}
\begin{center}
\includegraphics[width=\hsize,clip]{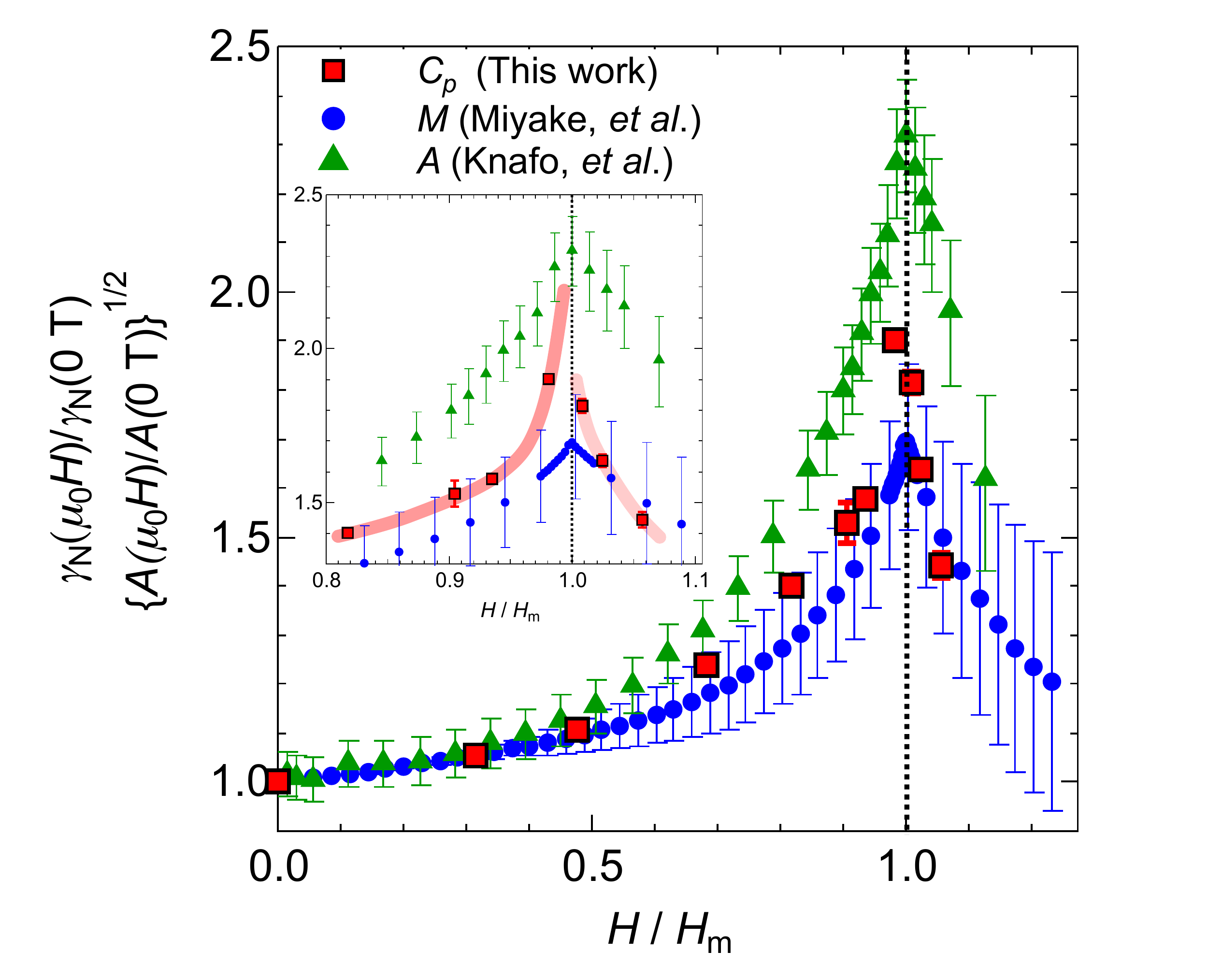}
\end{center}
\caption{(Color online)
Magnetic field dependence of $\gamma _{\rm N}$ with scaling by $\gamma _{\rm N}$ at 0~T and the metamagnetic transition field $H_{\rm m}$.
The red boxes and blue circles represent $\gamma$$_{\rm N}$($\mu$$_{0}$$H$)/$\gamma$$_{\rm N}$(0~T) measured in this work and estimated in the magnetization study~\cite{Miyake2019}, respectively.
The green triangles denote the square root of the normalized quadratic term $A$ obtained in the electrical resistivity measurement~\cite{Knafo2019}.
The inset shows the enlarged plot around $H_{\rm m}$ and the thick red curves are guides for the eyes.
}
\label{fig4}
\end{figure}

 In order to further explore the metamagnetic transition, we carried out the heat capacity measurements in pulsed magnetic fields up to 38.1 T, which are shown in Fig.~\ref{fig3}.
At 11.5~T, the higher temperature tail associated with the superconducting transition still remains below 1 K.
The persistence of the superconductivity against magnetic fields agrees with the exceptionally high upper critical fields~\cite{Ran2018,Aoki2019,Knebel2019,Ran2019},
in which the RSC up to $H_{\rm m}$ were reported for the perfect field-alignment along $b$-axis.
In the normal state, the intercepts of $C_p$/$T$ at $T = 0\,{\rm K}$ progressively increase as field increase and show a peak at $H_{\rm m}$, followed by the decrease at higher fields. 
Since the intercept of $C_p$/$T$ directly corresponds to $\gamma$$_{\rm N}$, this indicates that $\gamma$$_{\rm N}$ shows a peak at $H_{\rm m}$.
We should point out that the present $C_p$ data above $H_{\rm m}$ are missing below 1.5 K due to the large heating effect at $H_{\rm m}$ as shown in Fig.~\ref{fig2}.
One can still extrapolate the data and confirms that the $\gamma$$_{\rm N}$ decreases on the either side of $H_{\rm m}$.
In order to follow the field dependence of $\gamma$$_{\rm N}$ in more detail, we present the normalized electronic contribution $\gamma$$_{\rm N}$($\mu$$_{0}$$H$)/$\gamma$$_{\rm N}$(0~T) as a function of the normalized magnetic field $H$/$H_{\rm m}$ in Fig.~\ref{fig4}.
For comparison with earlier works, the normalized electronic contribution estimated by the magnetization measurement~\cite{Miyake2019}, 
and the square root of the normalized quadratic term $A$ of the electrical resistivity~\cite{Knafo2019} are also shown in Fig.~\ref{fig4}.
Here, the square root of $A$ is assumed to be proportional to the electronic heat capacity coefficient $\gamma$$_{\rm N}$ for a Fermi-liquid system, following the Kadowaki-Woods relation, which is often obeyed in heavy fermion compounds~\cite{Kad86}.
However, through ferromagnetic-paramagnetic instability crossing the Stoner factor $I=1$,
$A$ is predicted to diverge in $1/(1-I)$ while $\gamma$ varies as $\ln (1-I)$~\cite{xxx}.
Thus, the differences between $\sqrt{A}$ and $\gamma$ are expected. 
The $\gamma_{\rm N}$ rapidly increase with approaching $H_{\rm m}$, and the extrapolated value of $\gamma$$_{\rm N}$ to $H_{\rm m}$ is found to be more than twice ($\gamma$$_{\rm N}$($\mu$$_{0}$$H_{\rm m}$)~$\sim$250~mJK$^{-2}$mol$^{-1}$) at $H_{\rm m}$ than that at zero magnetic field.
The diverging behavior toward $H_{\rm m}$ qualitatively agrees with the earlier reports~\cite{Miyake2019,Knafo2019}.
We notice however there are a few quantitative differences; the asymmetry of $\gamma$$_{\rm N}$ around $H_{\rm m}$ and the absolute value of $\gamma$$_{\rm N}$($\mu$$_{0}$$H_{\rm m}$)/$\gamma$$_{\rm N}$(0~T).
The most plausible explanation for these discrepancies is the range of the measurement temperature.
This work estimates the density of state with the low temperature heat capacity from 0.8 to 2.5~K, while the resistivity (1.5 to 4.2~K) and magnetization (4.2 to 9.0~K) uses the data obtained relatively higher temperature region.
The earlier magnetization and resistivity measurements assume an isothermal condition for estimating the electronic density of state from field scan data.
The assumption might be broken with a rapid field sweep rate of pulsed fields, especially for UTe$_2$ due to the large temperature rise at $H_{\rm m}$ (See Fig.~\ref{fig2}).

 For the characterization of the metamagnetic transition, we roughly evaluate the latent heat and the hysteresis loss from the present results.
At the phase transition illustrated as the shaded area in Fig.~\ref{fig2}, we simply assume that the sample is in an adiabatic condition because the time-scale of the phase transition is about 10~msec that is much shorter than the external relaxation time ($>$100 msec).
In such cases, the entropy change $\Delta$$S$ can be expressed as the following formula~\cite{YAoki1998,Silhanek2006},
\begin{equation}
\Delta S=-C_p\Delta T/T+\delta Q_{\rm loss}/T,
\label{eq1}
\end{equation}
where $\delta Q_{\rm loss}$ represents the hysteresis loss.
This gives the estimates of the latent heat $\Delta S_{\rm m}$ and hysteresis loss $\delta$$Q_{\rm loss}$ as the half of the difference between the up-sweep and down-sweep results and the average of the up-sweep and down-sweep results, respectively.
Here, we assume the constant value of $C_p$/$T$=250~mJK$^{-2}$mol$^{-1}$ because $C_p$/$T$ is almost independent on temperature below 2~K even though the value should have some temperature, field dependences and be influenced by the difference between the up-sweep and down-sweep due to the hysteresis.
Since the ambiguity of the value is estimated as $\sim$10$\%$ by the discontinuity at $H_{\rm m}$ in the inset of Fig.~\ref{fig4}, our estimations should be accurate within the error margin of 10$\%$.
While the latent heat of the metamagnetic transition $\Delta S_{\rm m}$ is given as $\sim$$-$90~mJK$^{-1}$mol$^{-1}$, the $\delta Q_{\rm loss}$ reaches $\sim$320~mJmol$^{-1}$ when $T$ is about 1.5~K.
Using the relation of the hysteresis loss, $\delta$$Q_{\rm loss}$ = $\Delta M$ $\Delta H$, where $\Delta H$=0.1~T is the width of the hysteresis, the size of $\Delta M$ is given as $\Delta M$=0.6~$\mu _B$/f.u., which is well consistent with the reported magnetization jump~\cite{Miyake2019,Ran2019}.
Moreover, the $\Delta S_{\rm m}$ is also found to be fairly consistent with the value $-$70~mJK$^{-1}$mol$^{-1}$ calculated by using the magnetic Clausius-Clapeyron equation, $\Delta S_{\rm m}$=$-$$\Delta M$(d($\mu$$_{0}$$H_{\rm m})$/d$T$), with the slope of the reported phase boundary, d($\mu$$_{0}$$H_{\rm m})$/d$T$ of 20~mT/K~\cite{Miyake2019,Knafo2019} (the phase boundary is determined by the midpoints of the transition lines in the up-sweep and down-sweep processes), and the magnetization jump, $\Delta M$ of 0.6~$\mu _B$/f.u.
Although these analyses depend on the sample quality and measurement condition, the fulfillment of the relation confirms the reliability of our analysis and implies that the change in the entropy is predominately the magnetic origin.
If we take the latent heat as $\sim$$-$90~mJK$^{-1}$mol$^{-1}$, $\gamma$$_{\rm N}$ should show discontinuity with a step at $H_{\rm m}$, where the $\gamma$$_{\rm N}$ of the higher-field state ($>$$H_{\rm m}$) becomes smaller than that in the lower-field state ($<$$H_{\rm m}$).
As seen in the inset of Fig.~\ref{fig4}, our data are not sufficient to clearly see the discontinuity, but the $\gamma _{\rm N}$ of the higher-field state tends to be smaller than that of the lower-field state.
This finding that the electronic entropy below the critical field is higher than that above the critical field has an implication for the nature of the high field electronic state.
\begin{figure}
\begin{center}
\includegraphics[width=\hsize,clip]{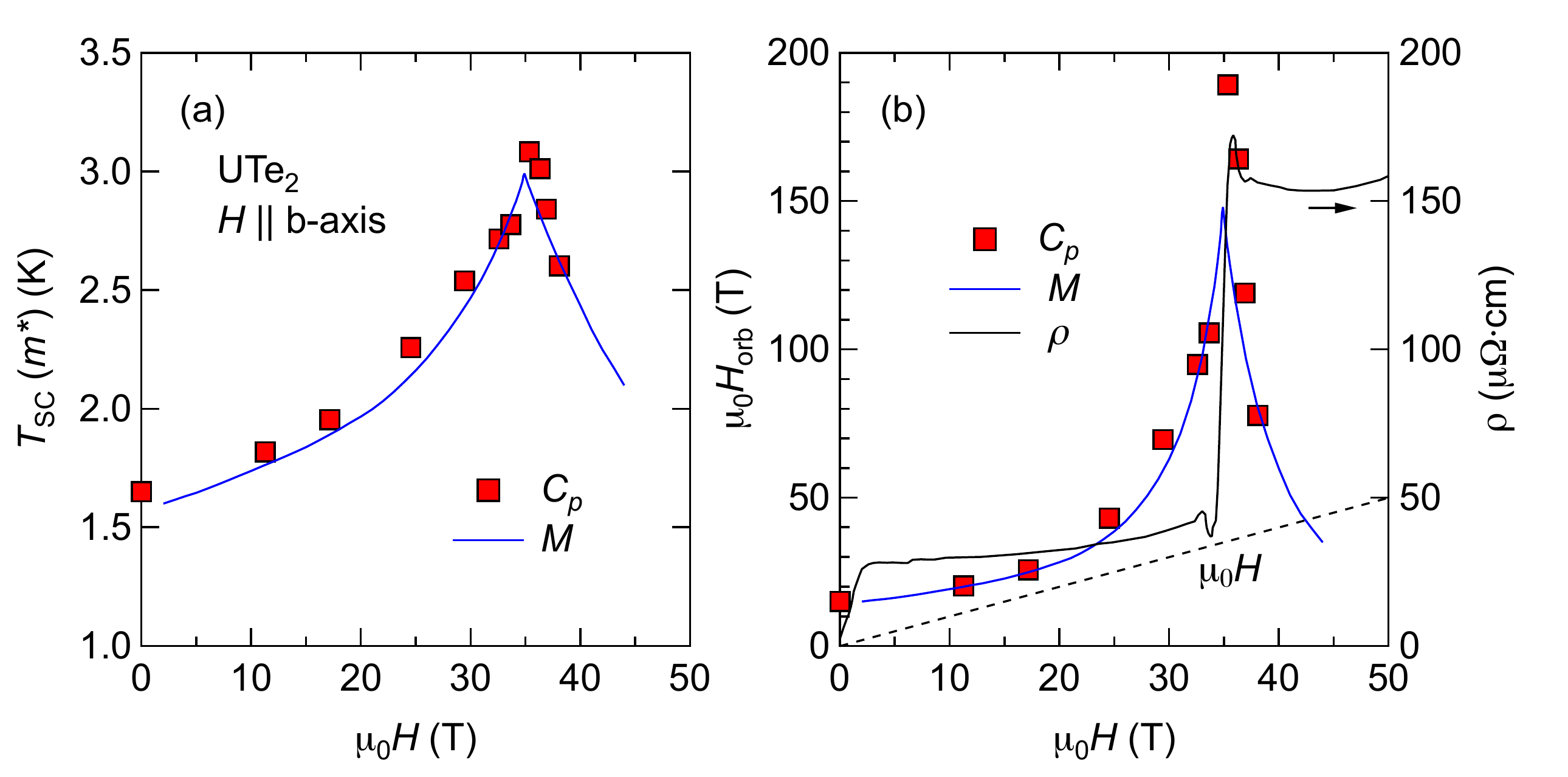}
\end{center}
\caption{(Color online) (a) Calculated $T_{\rm SC}$ as a function of field obtained from the heat capacity and magnetization~\cite{Miyake2019} measurements in UTe$_2$. 
(b) Calculated orbital limits (left-axis) and electrical resistivity~\cite{Knafo2019} (right-axis) as a function of field. The orbital limits are obtained by the formula $H_{\rm orb}\sim (m^\ast T_{\rm SC})^2$ from the data of the heat capacity and magnetization~\cite{Miyake2019} measurements.
}
\label{fig5}
\end{figure}
Figure~\ref{fig5}(a) shows the field dependence of $T_{\rm SC}$ derived from
heat capacity and magnetization data by using a simplified McMillan-type formula~\cite{Aok19,Miy08,Miy09,Miyake2019}, $T_{\rm SC}=T_0\exp\left({-\frac{\lambda + 1}{\lambda}}\right)$, where $T_0$ is a constant determined by the experimental $T_{\rm SC}$ at 0~T, with a parameter chosen in Ref.~\citen{Miyake2019},
assuming $m_{\rm b}$ is field-independent.
Figure~\ref{fig5}(b) emphasizes the huge enhancement of the orbital limit $H_{\rm orb}$ by a factor $\sim 13$ close to $H_{\rm m}$, assuming $H_{\rm orb}$ proportional to $(m^\ast T_{\rm SC})^2$.
Let us remark that $H_{\rm c2}$ becomes closer to the applied field $H$ in the field range between $12$ and $25\,{\rm T}$, where RSC is observed for the perfect $H$ alignment along $b$-axis.
For $H\parallel b$, superconductivity disappears above $H_{\rm m}$.
Two main mechanisms are:
i) change of Fermi surface with an enhancement of $m_{\rm b}$ and thus a drop of $\lambda$ above $H_{\rm m}$, 
ii) drastic decrease of the mean free path $l$ above $H_{\rm m}$, leading to the collapse of superconductivity.
A remarkable phenomenon is the jump of the magnetoresistance at $1.5\,{\rm K}$ just
on entering into the polarized ferromagnetic phase~\cite{Knebel2019}.
If we assume that $\rho$ just above $H_{\rm m}$ is related to the electronic disorder going from PM to FM ground state,
$T_{\rm SC}(m^\ast_H)/T_{\rm SC}(m^\ast_{H=0})$ is boosted by a factor 2,
while $\rho$ is jumped by a factor 5.
It is worthwhile to remember the difficulty to evaluate $A$ in a clean material
when the crossover from collision regime to collisionless regime at $\omega _{\rm c}\tau$ reaches $1$,
where $\omega _{\rm c}$ ($=e$($\mu$$_{0}$$H)/(m^\ast c)$) is cyclotron frequency, $\tau$ ($\propto l$) is the scattering life time.
Fortunately just around $H_{\rm m}$, $\omega_{\rm c}\tau$ drops by a factor near 10 due to the increase of $m^\ast$
and the decrease of $l$.
In order to clarify the angular dependence of superconducting stability in the PM and polarized FM phases,
important ingredients are the angular dependence of $m^\ast$($\mu$$_{0}$$H)$, $m_{\rm b}$($\mu$$_{0}$$H)$ and $l$($\mu$$_{0}$$H)$.
Note that it is established that field misalignment to $a$-axis leads to a fast increase of $H_{\rm m}$ associated to a collapse of RSC,
as it occurs in URhGe~\cite{Lev05},
while misalignment to $c$-axis leads to a weak increase of $H_{\rm m}$ and 
stabilization of a superconducting ferromagnetic domain as observed in UGe$_2$~\cite{She01}.
From previous studies on heavy fermion compounds~\cite{Flo88} and extensive measurements on the link between $m^\ast$($\mu$$_{0}$$H_{\rm m})$ and $\rho$($\mu$$_{0}$$H_{\rm m})$ in URhGe~\cite{Lev05,Lev07,Miy08,Miy09,Har11,Nak17},
the reappearance of SC in FM phase may be explained.

 In summary, we studied the thermodynamic properties of the heavy fermion superconductor UTe$_2$ in pulsed high magnetic fields applied parallel to the $b$-axis so as to elucidate the details of the recently reported metamagnetic transition~\cite{Miyake2019,Knafo2019,Ran2019}.
As reported in the previous works~\cite{Ran2018,Aoki2019}, we confirm the large electronic heat capacity coefficient of the normal state, $\gamma$$_{\rm N}$= ~123.5$\pm$0.3 mJK$^{-2}$mol$^{-1}$, at 0 T.
The sharp heat capacity jump by the superconducting transition at $T_{\rm c}{\sim}$1.65~K is also reproduced.
From the results of the magnetocaloric effect in pulsed-fields, we detect the thermodynamic anomaly at $\sim$36.0~T.
The heat capacity measurements in pulsed high fields reveal the diverging behavior of $\gamma$$_{\rm N}$ toward $H_{\rm m}$, although the superconducting transition cannot be detected above 11.5~T due to the slight tilt of the $b$-axis from the field direction.
Using the present results, the details of the metamagnetic transition is clarified as a first-order transition with the latent heat and hysteresis loss and we quantitatively succeed to demonstrate the magnetization work~\cite{Miyake2019,Ran2019} through the magnetic Clausius-Clapeyron relation.
The significant development of $\gamma$$_{\rm N}$ directly indicates the enhancement of $m^\ast$ linked to FM instability driven at $H_{\rm m}$.
Qualitatively the RSC can be well explained by the field dependence of $m^\ast$ and the feedback on the electronic mean free path.
As the future issues, the pairing mechanism of the superconductivity including symmetry of the gap function should be investigated.
A clear continuation of our thermodynamic studies is to detect the $m^\ast$($\mu$$_{0}$$H)$
singularities in the SC-FM domain detected with a misalignment along $c$-axis.

\begin{acknowledgment}
This work was supported by  KAKENHI (JP15H05884, JP15H05882, JP15K21732, JP16H04006, JP15H05745).
\end{acknowledgment}

%
%
%

\renewcommand{\thefigure}{S\arabic{figure}}
\clearpage
\begin{center}
\large{\bf{Supplemental Material for\\
Thermodynamic Investigation of Metamagnetism in Pulsed High Magnetic Fields on Heavy Fermion Superconductor UTe$_2$
}}
\end{center}

\begin{figure}[b]
\begin{center}
\includegraphics[width=\hsize,clip]{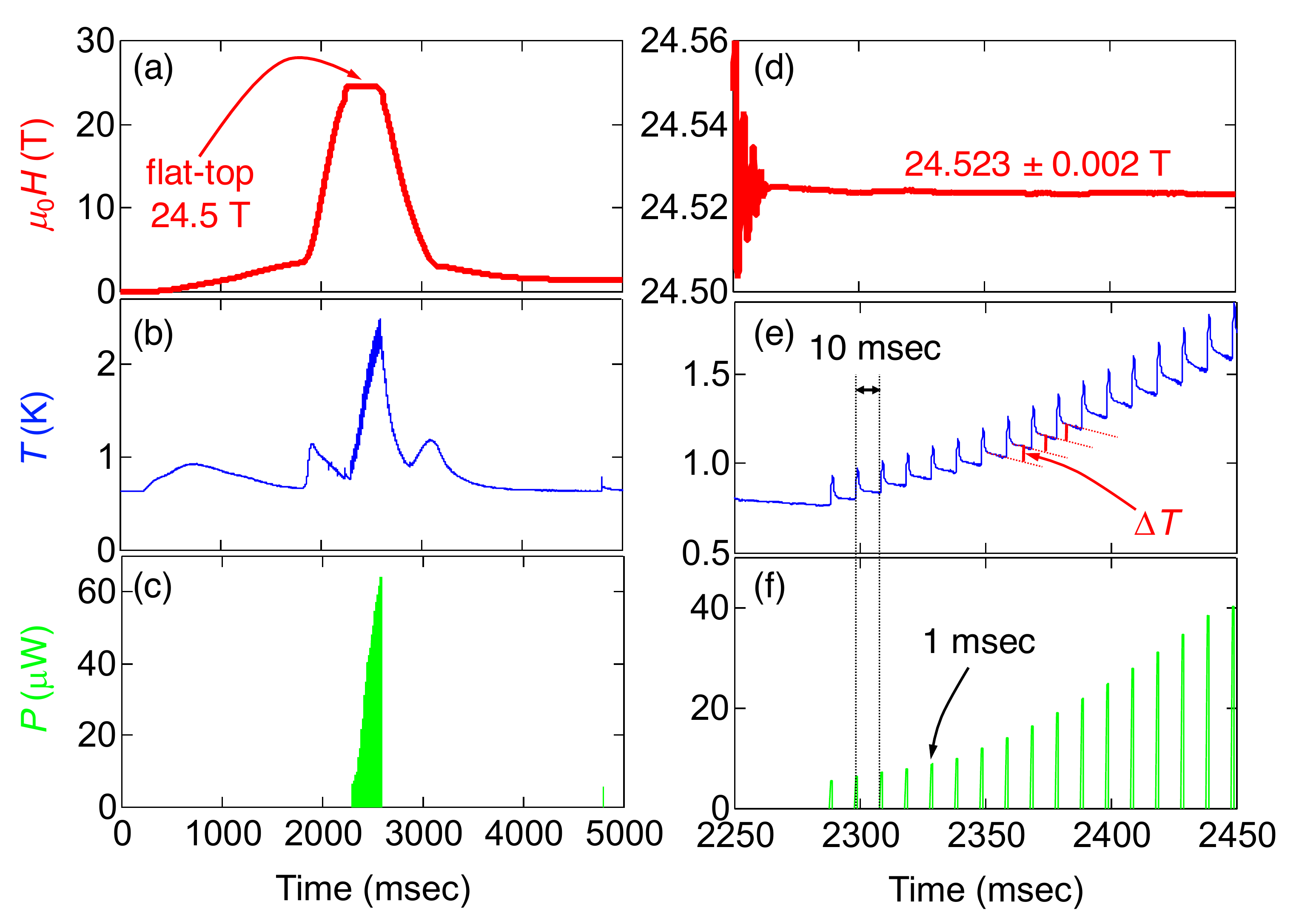}
\end{center}
\caption{(Color online) Quasi-adiabatic heat capacity experiment in the long pulsed magnet. (a) Time dependence of magnetic field profile, (b) Sample temperature, (c) applied power as a function of time.
(d),(e),(f) are the enlarged plots of (a),(b),(c), respectively, for the stabilized field region from 2250 to 2450 ms.
In each 10 msec temperature step, heat capacity is measured by the simple equation $C_p$=$\Delta Q$/$\Delta T$, where $\Delta Q$ is a total of the applied heat quantity in each heat pulse.}
\label{figS1}
\end{figure}

\vspace{\baselineskip}
In this supplemental material, the details of heat capacity measurements under pulsed magnetic fields are presented.
Measurements of heat capacity in pulsed magnetic fields were performed in highly stabilized magnetic fields generated by the specially designed long pulsed magnet.
The pulsed magnet was composed of a main coil and an additional mini-coil to generate highly stabilized magnetic fields for several hundreds miliseconds with a feedback control technique~\cite{Kohama2015}.
Fig.~\ref{figS1}(a),(d) show a typical magnetic field profile used for this research, where the total time duration of pulse is $\sim$1.3 sec.
The field stability during heat capacity measurements on the top of field pulse is within $\sim$0.002 T.
To obtain the absolute value of heat capacity, we employed the quasi-adiabatic method~\cite{Kohama2013,Kohama2019,Jiao2019}.
The calorimeter consists of liquid $^3$He thermal bath, Pt$_{0.92}$W$_{0.08}$ thermal relaxation wires and TiO$_2$ substrate to which RuO$_2$ resistors for a heater and a thermometer are attached. 
The single crystal of UTe$_2$ was mounted on the substrate of the calorimeter with Apiezon N grease.

As seen in Fig.~\ref{figS1}(d-f), the heat capacity measurements were carried out within the stabilized field region, where each data point was obtained in 10 msec with a 1 msec heat pulse.
The duration of measurement time scale of 10~ms was enough shorter than the external relaxation time $\tau _1$ that was the order of 100~msec.
The internal relaxation time of the sample $\tau _2$ was about one millisecond as seen in the rapid thermal relaxation after the application of the heat pulse (see Fig.~\ref{figS1}(e)).
We can avoid the effect of $\tau _2$ by using the data after the completion of the internal relaxation as illustrated by the red dotted lines.
Therefore, the present condition, $\tau _2\ll 10~{\rm msec}\ll\tau _1$, is suitable for the quasi-adiabatic method and the heat capacity is simply given as $C_p$=$\Delta Q$/$\Delta T$ where $\Delta Q$ and $\Delta T$ are total amount of applied heat and temperature increment, respectively.


\begin{thebibliography}{99}
\bibitem{Sax00} S.S. Saxena, P. Agarwal, K. Ahilan, F.M. Grosche, R.K.W. Haselwimmer, M.J. Steiner, E. Pugh, I.R. Walker, S.R. Julian, P. Monthoux, G.G. Lonzarich, A. Huxley, I. Sheikin, D. Braithwaite and J. Flouquet, Nature {\bf 406}, 587 (2000).
\bibitem{Aok19} D. Aoki, K. Ishida and J. Flouquet, J. Phys. Soc. Jpn. {\bf 88}, 022001 (2019).
\bibitem{Aok12_JPSJ_review} D. Aoki and J. Flouquet, J. Phys. Soc. Jpn. {\bf 81}, 011003 (2012).
\bibitem{Aok01} D. Aoki, A. Huxley, E. Ressouche, D. Braithwaite, J. Flouquet, J.-P. Brison, E. Lhotel and C. Paulsen, Nature {\bf 413}, 613 (2001).
\bibitem{Lev05} F. L\'{e}vy, I. Sheikin, B. Grenier and A.D. Huxley, Science {\bf 309}, 1343 (2005).
\bibitem{Lev07} F. L\'{e}vy, I. Sheikin and A. Huxley, Nature Physics {\bf 3}, 460 (2007).
\bibitem{Miy08} A. Miyake, D. Aoki and J. Flouquet, J. Phys. Soc. Jpn. {\bf 77}, 094709 (2008).
\bibitem{Miy09} A. Miyake, D. Aoki and J. Flouquet, J. Phys. Soc. Jpn. {\bf 78}, 063703 (2009).
\bibitem{She18} Y. Sherkunov, A.V. Chubukov and J.J. Betouras, Phys. Rev. Lett. {\bf 121}, 097001 (2018).
\bibitem{Tau10} V. Taufour, D. Aoki, G. Knebel and J. Flouquet, Phys. Rev. Lett. {\bf 105}, 217201 (2010).
\bibitem{Kot11} H. Kotegawa, V. Taufour, D. Aoki, G. Knebel and J. Flouquet, J. Phys. Soc. Jpn. {\bf 80}, 083703 (2011).
\bibitem{Aok11_UCoAl} D. Aoki, T. Combier, V. Taufour, T.D. Matsuda, G. Knebel, H. Kotegawa and J. Flouquet, J. Phys. Soc. Jpn. {\bf 80}, 094711 (2011).
\bibitem{Kim15} N. Kimura, N. Kabeya, H. Aoki, K. Ohyama, M. Maeda, H. Fujii, M. Kogure, T. Asai, T. Komatsubara, T. Yamamura and I. Satoh, Phys. Rev. B {\bf 92}, 035106 (2015).
\bibitem{Ran2018} S. Ran, C. Eckberg, Q.-P. Ding, Y. Furukawa, T. Metz, S. R. Saha, I. -L. Liu, M. Zic, H. Kim, J. Paglione, and N. P. Butch, arXiv:1811.11808.
\bibitem{Aoki2019} D. Aoki, A. Nakamura, F. Honda, D. X. Li, Y. Homma, Y. Shimizu, Y. J. Sato, G. Knebel, J.-P. Brison, A. Pourret, D. Braithwaite, G. Lapertot, Q. Niu, M. Vali$\check{\rm s}$ka, H. Harima, and J. Flouquet, J. Phys. Soc. Jpn. {\bf 88}, 043702 (2019).
\bibitem{Sundar2019} S. Sundar, S. Gheidi, K. Akintola, A. M. C$\hat{\rm o}$t$\acute{\rm e}$, S. R. Dunsiger, S. Ran, N. P. Butch, S. R. Saha, J. Paglione, and J. E. Sonier, arXiv:1905.06901.
\bibitem{Miyake2019} A. Miyake, Y. Shimizu, Y. J. Sato, D.X. Li, A. Nakamura, Y. Homma, F. Honda, J. Flouquet, M. Tokunaga, and D. Aoki, J. Phys. Soc. Jpn.  {\bf 88}, 063706 (2019).
\bibitem{Knafo2019} W. Knafo, M. Vali$\check{\rm s}$ka, D. Braithwaite, G. Lapertot, G. Knebel, A. Pourret, J.-P. Brison, J. Flouquet, D. Aoki, J. Phys. Soc. Jpn. {\bf 88}, 063705 (2019).
\bibitem{Ran2019} S. Ran ,I-L. Liu, Y. S. Eo, D. J. Campbell, P. Neves, W. T. Fuhrman, S. R. Saha, C. Eckberg, H. Kim, J. Paglione, D. Graf, J. Singleton and N. P. Butch, arXiv:1905.04343.
\bibitem{Knebel2019} G. Knebel, W. Knafo, A. Pourret, Q. Niu, M. Vališka, D. Braithwaite, G. Lapertot, M. Nardone, A. Zitouni, S. Mishra, I. Sheikin, G. Seyfarth, J.-P. Brison, D. Aoki, and J. Flouquet, J. Phys. Soc. Jpn. {\bf 88}, 063707 (2019).
\bibitem{Kohama2013} Y. Kohama, Y. Hashimoto, S. Katsumoto, M. Tokunaga and K. Kindo,  Meas. Sci. Technol. {\bf 24}, 115005 (2013).
\bibitem{Kohama2019} Y. Kohama, H. Ishikawa, A. Matsuo, K. Kindo, N. Shannon, and Z. Hiroi, Proc. Natl. Acad. Sci. USA, {\bf 116} 10686 (2019).
\bibitem{Jiao2019} L. Jiao, M. Smidman, Y. Kohama, Z. S. Wang, D. Graf, Z. F. Weng, Y. J. Zhang, A. Matsuo, E. D. Bauer, Hanoh Lee, S. Kirchner, J. Singleton, K. Kindo, J. Wosnitza, F. Steglich, J. D. Thompson, and H. Q. Yuan, Phys. Rev. B {\bf 99}, 045127 (2019).
\bibitem{Kohama2015} Y. Kohama, and K. Kindo,  Rev. Sci. Instrum. {\bf 86}, 104701 (2015).
\bibitem{suppl} (Supplemental Material) The details of the present heat capacity measurements are provided online.
\bibitem{Silhanek2006} A. V. Silhanek, M. Jaime, N. Harrison, V. R. Fanelli, C. D. Batista, H. Amitsuka, S. Nakatsuji, L. Balicas, K. H. Kim, Z. Fisk, J. L. Sarrao, L. Civale, and J. A. Mydosh, Phys. Rev. Lett. {\bf 96}, 136403 (2006).
\bibitem{Kad86} K. Kadowaki and S. B. Woods, Solid State Commun. {\bf 58}, 507 (1986).
\bibitem{xxx} T. Moriya, Acta Phys. Pol. B {\bf 34}, 287 (2003).
\bibitem{YAoki1998} Y. Aoki, T. D.Matsuda, H. Sugawara, H. Sato, H. Ohkuni, R. Settai, Y. $\bar{\rm O}$nuki, E. Yamamoto, Y. Haga, A. V. Andreev, V. Sechovsky, L. Havela, H. Ikeda, K. Miyake, J. Magn. Magn. Mater. {\bf 177-181}, 271 (1998).
\bibitem{She01} I. Sheikin, A. Huxley, D. Braithwaite, J.P. Brison, S. Watanabe, K. Miyake and J. Flouquet, Phys. Rev. B {\bf 64}, 220503 (2001).
\bibitem{Flo88} J. Flouquet, P. Haen, F. Lapierre, C. Fierz, A. Amato and D. Jaccard, J. Magn. Magn. Mater. {\bf 76{\&}77}, 285 (1988).
\bibitem{Har11} F. Hardy, D. Aoki, C. Meingast, P. Schweiss, P. Burger, H. v. Loehneysen and J. Flouquet, Phys. Rev. B {\bf 83}, 195107 (2011).
\bibitem{Nak17} S. Nakamura, T. Sakakibara, Y. Shimizu, S. Kittaka, Y. Kono, Y. Haga, J. Posp{\'i}{\v{s}}il and E. Yamamoto, Phys. Rev. B {\bf 96}, 094411 (2017).
\end{thebibliography}

\begin{thebibliography}{99}
%
\bibitem{Kohama2015} Y. Kohama, and K. Kindo,  Rev. Sci. Instrum. {\bf 86}, 104701 (2015).
%
\bibitem{Kohama2013} Y. Kohama, Y. Hashimoto, S. Katsumoto, M. Tokunaga and K. Kindo,  Meas. Sci. Technol. {\bf 24}, 115005 (2013).
%
\bibitem{Kohama2019} Y. Kohama, H. Ishikawa, A. Matsuo, K. Kindo, N. Shannon, and Z. Hiroi, Proc. Natl. Acad. Sci. USA, {\bf 116}, 10686 (2019).
%
\bibitem{Jiao2019} L. Jiao, M. Smidman, Y. Kohama, Z. S. Wang, D. Graf, Z. F. Weng, Y. J. Zhang, A. Matsuo, E. D. Bauer, Hanoh Lee, S. Kirchner, J. Singleton, K. Kindo, J. Wosnitza, F. Steglich, J. D. Thompson, and H. Q. Yuan, Phys. Rev. B {\bf 99}, 045127 (2019).

\end{thebibliography}
\end{document}